\documentclass[preprint,sts]{imsart}

\usepackage[dvips]{graphicx}
\usepackage{amssymb,amsfonts,amsmath,amsthm,color}

\newtheorem{theorem}{Theorem}
\newtheorem{lemma}{Lemma}

\newcommand{\RR}{\mathbb{R}}
\newcommand{\R}{\mathbb{R}}
\newcommand{\ZZ}{\mathbb{Z}}

\newcommand{\E}{\mathbb{E}}

\newcommand{\NNN}{\mathcal{N}}
\newcommand{\BBB}{\mathcal{B}}

\newcommand{\cN}{\mathcal{N}}
\newcommand{\metric}{\rho}

\newcommand{\sym}{\operatorname{sym}}
\newcommand{\poly}{\operatorname{poly}}

\newcommand{\eps}{\varepsilon}
\newcommand{\one}{\mathbf{1}}
\newcommand{\SNR}{\mathrm{SNR}}
\newcommand{\cS}{\mathcal{S}}

\newcommand{\epr}{\hfill$\Box$\linebreak}

\begin{document}
\begin{frontmatter}

\title{The sample complexity of multi-reference alignment}
\runtitle{Sample complexity of MRA}

    \begin{aug}
\author{\fnms{Amelia}~\snm{Perry}\thanksref{t1}\ead[label=amelia]{ameliaperry@mit.edu}},
\author{\fnms{Jonathan}~\snm{Weed}\thanksref{t2}\ead[label=jweed]{jweed@mit.edu}},
\author{\fnms{Afonso S.}~\snm{Bandeira}\thanksref{t3}\ead[label=afonso]{bandeira@cims.nyu.edu}},
\author{\fnms{Philippe}~\snm{Rigollet}\thanksref{t4}\ead[label=rigollet]{rigollet@math.mit.edu}}
\and
\author{\fnms{Amit}~\snm{Singer}\thanksref{t5}\ead[label=amit]{amits@princeton.edu}}

\affiliation{Massachusetts Institute of Technology\\
Massachusetts Institute of Technology\\
Courant Institute of Mathematical Sciences, New York University\\
Massachusetts Institute of Technology\\
Princeton University}
\thankstext{t1}{Supported in part by NSF grant DMS-1541100, NSF CAREER Award CCF-1453261 and a grant from the MIT NEC Corporation.}
\thankstext{t2}{Supported in part by NSF Graduate Research Fellowship DGE-1122374.}
\thankstext{t3}{Supported in part by NSF grants DMS-1317308, DMS-1712730, and DMS-1719545. Part of this work was done while ASB was with the Mathematics Department at MIT.}
\thankstext{t4}{Supported in part by NSF grants CAREER DMS-1541099, DMS-1541100, DMS-1712596 and DMS-TRIPODS-1740751; DARPA grant W911NF-16-1-0551, ONR grant N00014-17-1-2147, a grant from the MIT NEC Corporation, grant 2018-182642 from the Chan Zuckerberg Initiative DAF and the MIT Skoltech Seed Fund.
}
\thankstext{t5}{Supported in part by ARO grant W911NF-17-1-0512, Award Number R01GM090200 from the NIGMS, FA9550-17-1-0291 from AFOSR, Simons Investigator Award and Simons Collaboration on Algorithms and Geometry from Simons Foundation, and the Moore Foundation Data-Driven Discovery Investigator Award.
}

\address{{Amelia Perry}\\
{Department of Mathematics} \\
{Massachusetts Institute of Technology}\\
{77 Massachusetts Avenue,}\\
{Cambridge, MA 02139-4307, USA}\\
\printead{amelia}
}

\address{{Jonathan Weed}\\
{Department of Mathematics} \\
{Massachusetts Institute of Technology}\\
{77 Massachusetts Avenue,}\\
{Cambridge, MA 02139-4307, USA}\\
\printead{jweed}
}

\address{{Afonso S.\ Bandeira}\\
{Department of Mathematics} \\
{Courant Institute of Mathematical Sciences}\\
{Center for Data Science} \\
{New York University,}\\
{New York, NY 10012, USA}\\
\printead{afonso}
}

\address{{Philippe Rigollet}\\
{Department of Mathematics} \\
{Massachusetts Institute of Technology}\\
{77 Massachusetts Avenue,}\\
{Cambridge, MA 02139-4307, USA}\\
\printead{rigollet}
}

\address{{Amit Singer}\\
{Department of Mathematics} \\
{Program in Applied and Computational Mathematics}\\
{Princeton University,}\\
{Princeton NJ, 08544, USA}\\
\printead{amit}
}

\runauthor{Perry et al.}
\end{aug}
\def\abstractname{Abstract}

\begin{abstract}
The growing role of data-driven approaches to scientific discovery has unveiled a large class of models that involve latent transformations with a rigid algebraic constraint. Three-dimensional molecule reconstruction in Cryo-Electron Microscopy (cryo-EM) is a central problem in this class.
Despite decades of algorithmic and software development, there is still little theoretical understanding of the sample complexity of this problem, that is, number of images required for 3-D reconstruction. Here we consider {multi-reference alignment} (MRA), a simple model that captures fundamental aspects of the statistical and algorithmic challenges arising in cryo-EM and related problems. In MRA, an unknown signal is subject to two types of corruption: a latent cyclic shift and the more traditional additive white noise. The goal is to recover the signal at a certain precision from independent samples.  
While at high signal-to-noise ratio (SNR), the number of observations needed to recover a generic signal is proportional to $\mathbf{1/\text{{\bf SNR}}}$, we prove that it rises to a surprising $\mathbf{1/\text{{\bf SNR}}^3}$ in the low SNR regime. 
This precise phenomenon was observed empirically more than twenty years ago for cryo-EM but has remained unexplained to date.
Furthermore, our techniques can easily be extended to the heterogeneous MRA model where the samples come from a mixture of signals, as is often the case in applications such as cryo-EM, where molecules may have different conformations. This provides a first step towards a statistical theory for heterogeneous cryo-EM.
\end{abstract}

\begin{keyword}[class=AMS]
\kwd[Primary ]{62B10}
\kwd[; secondary ]{92C55}
\end{keyword}
\begin{keyword}[class=KWD]
Multi-reference alignment, method of invariants, bispectrum, cryo-EM, tensor decomposition
\end{keyword}
\end{frontmatter}

\section{Introduction}\label{sec:intro}

Sample complexity is a concept at the cornerstone of statistics and machine learning with far reaching implications for experimental design and data collection strategies, ranging from polling voters for election prediction to training speech recognition systems. 
Loosely speaking, the sample complexity is the number of measurements needed to estimate model parameters at a prescribed accuracy. Perhaps the most fundamental question associated to sample complexity is its scaling with respect to signal-to-noise ratio (SNR) of the problem at hand. This question is of prime importance especially in modern problems arising in data-driven science,  which often feature a very low SNR.

In many traditional models, the sample complexity scales as 1/SNR, but it is significantly more difficult to establish analogous results for more complex models, such as those which feature latent variables in order to account for heterogeneity in the data. In this paper,  we examine the sample complexity of complex models in which the signal undergoes two types of corruption: a latent linear transformation and noise addition. Of particular interest in applications are linear transformations that correspond to a group action. For example, in estimating a two-dimensional image from multiple arbitrarily rotated noisy copies,
every measurement corresponds to an unknown element of the group of planar rotations $\mathrm{SO}(2)$ that acts linearly on the data. Another example is the reconstruction problem in cryo-EM~\cite{Fra06}, a fundamental imaging technique that won the 2017 Nobel Prize in Chemistry. In cryo-EM, the goal is to estimate the three-dimensional structure of a molecule from many two-dimensional noisy tomographic projection images taken at unknown viewing angles.
Here to every projection image corresponds an unknown element of the 3D rotation group $\mathrm{SO}(3)$ and the linear transformation is a composition of a tomographic projection in a fixed direction with the group action of rotating the molecular structure (we ignore possible in-plane translations and other imaging effects). Other estimation problems of similar nature arise in many other scientific and engineering disciplines, such as structure from motion (SfM) in computer vision~\cite{SAgarwal_etal_2009_Rome}, simultaneous localization and mapping (SLAM) in robotics~\cite{Rosen_SLAM_PCC}, X-ray free electron lasers (XFEL) in structural biology~\cite{ChaBarBog06,GafCha07},
crystalline simulations~\cite{sonday2011_sim}, and shape matching and image registration and alignment problems arising in geology, medicine, and paleontology, to name a few~\cite{dryden98_shape,foroosh02_subpixelregistration,SchWanZha18}.

Multi-reference alignment (MRA)~\cite{Bandeira_Charikar_Singer_Zhu_Alignment} is one of the simplest models that is able to capture fundamental aspects of this class of problems, rendering it ideal for theoretical study. In this model one observes $n$ independent data points $y_1, \dots, y_n$ given by
\begin{equation}\label{eq:MRA:model}
y_i = R_{\ell_i}\theta + \sigma \xi_i,
\end{equation}
where   $R_{\ell_i}$ is a cyclic shift by an unknown number $\ell_i$ of coordinates: the $j$th coordinate of $R_{\ell_i}\theta \in \R^d$ is given by $\big(R_{\ell_i}\theta)_j=\theta_{j+\ell_i \pmod{d}}$. We assume isotropic Gaussian noise $\xi_i \sim\NNN(0,I_{d})$ i.i.d.\ and independent of $\ell_1, \dots, \ell_n$. We make no assumptions on the shifts $\ell_1, \dots, \ell_n$.
However, we can always reduce to the case where $\ell_1, \ldots, \ell_n$ are drawn i.i.d.\ uniformly from $[d]$: indeed, if we let $\ell'_1, \dots, \ell'_n$ be i.i.d.\ uniform from $[d]$ and independent of all other random variables, the sums $\ell_i + \ell'_i \pmod{d}$ for $i = 1, \dots, n$ are i.i.d.\ uniform in $[d]$.
Since the Gaussian distribution is invariant under cyclic shifts, we can therefore replace the observation $y_i$ by $R_{\ell'_i}y_i = R_{\ell'_i+ \ell_i} \theta + \sigma R_{\ell'_i} \xi_i$ and reduce to the MRA model where the shifts are assumed to be uniform.
We therefore focus on this case for simplicity and generality. The goal is to estimate the unknown vector $\theta \in \RR^d$.

\begin{figure}
\begin{center}
\includegraphics[width=0.8\textwidth]{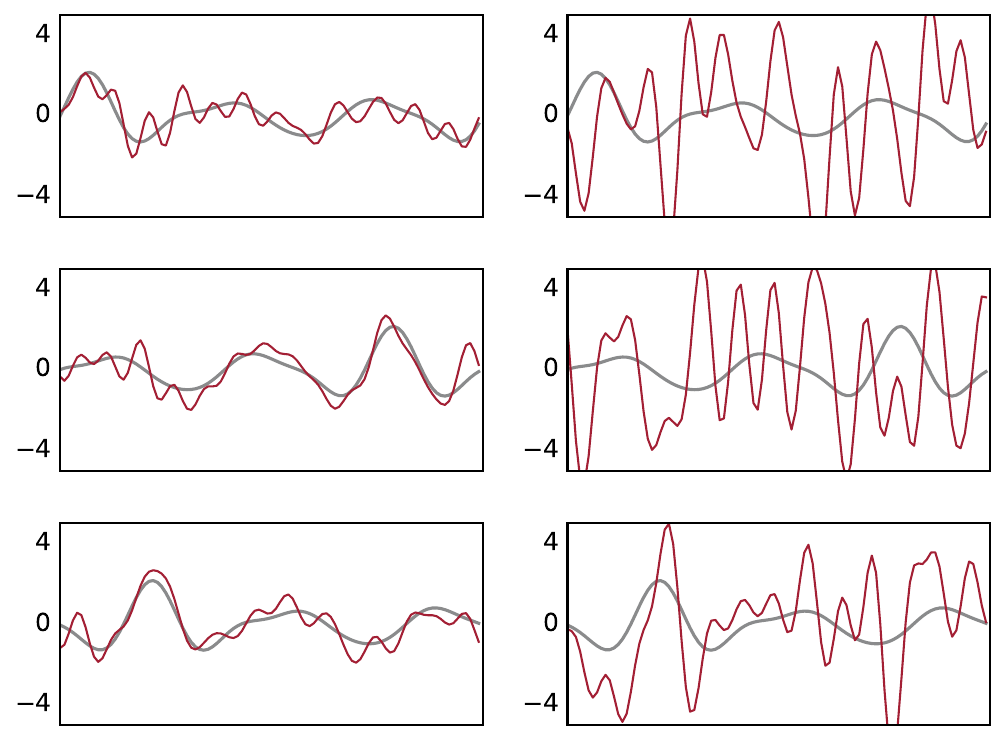}
\caption{Instances of the multi-reference alignment problem, at low ($\sigma \approx .5$, left column) and high ($\sigma \approx 3$, right column) noise levels. We plot the values of a vector in $\RR^d$ for $d = 100$. Randomly shifted copies of a smoothed version of the underlying signal $(\theta)$ appears in gray, and a smoothed version of the noisy observation $(y)$ appears in red. When the noise level is low, salient features of the signal are still visible despite the noise; in the presence of large noise, however, the signals cannot reliably be aligned. We establish the optimal sample complexity of the large noise problem.\label{fig:noise_low_high}}
\end{center}
\end{figure}

The MRA model is illustrated in Figure~\ref{fig:noise_low_high}.  
We refer to $\|\theta\|_2^2/\sigma^2$ as the $\SNR$; without loss of generality we assume in the sequel that $\|\theta\|_2=1$, implying $\SNR=1/{\sigma^2}$. 
The latent transformations $R_\ell$ in MRA correspond to the action of the cyclic group $\mathbb{Z}/d\mathbb{Z}$ on real-valued signals of length $d$. The simplicity of MRA in the class of problems mentioned earlier stems from the following facts: (i) the group $\mathbb{Z}/d\mathbb{Z}$ is finite (has exactly $d$ elements) and commutative (i.e., $R_\ell R_m = R_m R_\ell$ for all $\ell,m$), and (ii) no further linear operation (such as projection as in cryo-EM) is involved.

\begin{figure}
\begin{center}
\includegraphics[width=0.65\textwidth]{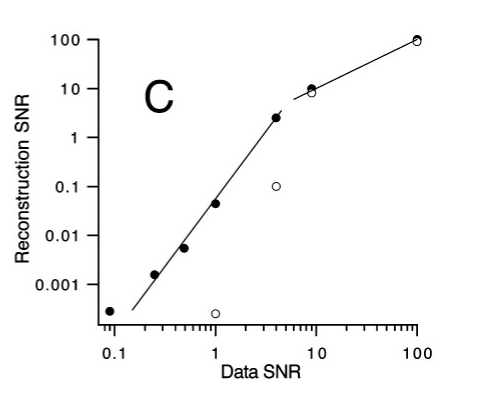}
\caption{Figure taken from a paper on cryo-EM~\cite{Sig98} which examined the empirical performance of maximum-likelihood estimation for two-dimensional MRA, illustrating (i) strikingly different behavior for the maximum-likelihood estimator in low and high SNR regimes and (ii) $1/\mathrm{SNR}^3$ scaling at low SNR. Our theoretical analysis suggests that any estimator, not only the maximum-likelihood estimator, is bound to the same limitations. Reprinted from \emph{Journal of Structural Biology}, Vol.\ 122, F.\ Sigworth, A maximum-likelihood approach to single-particle image refinement, pp.\ 328--339, copyright 1998, used with permission. \label{fig:fred}}
\end{center}
\end{figure}

In this paper, we study the sample complexity of MRA, that is, the number of observations needed to recover a generic signal with a given accuracy as a function of the $\SNR$.
Our results reveal a striking difference between the high and low $\SNR$ regimes. On the one hand, the picture at a high $\SNR$ is fairly standard in signal processing: the sample complexity scales proportional to $1/\mathrm{SNR}$. On the other hand, using information theoretic arguments, we show that the presence of the latent cyclic shifts has a profound effect on the sample complexity at low $\SNR$, where the optimal sample complexity becomes proportional to  $1/\mathrm{SNR}^3$.
Twenty years ago, in a seminal paper by Sigworth~\cite{Sig98} that introduced maximum-likelihood estimation to the cryo-EM field, an analogous phenomenon was empirically observed (without theoretical explanation) in two-dimensional multi-reference alignment (Figure~\ref{fig:fred}), where the group of transformations are planar rigid motions.
Our results shed light on the fundamental reasons behind this behavior of the sample complexity.

More specifically, our results on the sample complexity of MRA highlight the role of the \emph{third moment tensor} (known in signal processing as the \emph{bispectrum}) in the estimation task. From this analysis, we not only show that the  $1/\mathrm{SNR}^3$ dependence is unavoidable for \emph{any} method in the low-SNR regime, but also give a very simple algorithm based on tensor decomposition, which achieves the optimal sample complexity efficiently and provably. 

By establishing the correct sample complexity for the MRA model, this work represents the first step towards determining the sample complexity of the reconstruction problem in cryo-EM and other applications involving more complicated group actions. In fact, we complement our results on MRA by showing that a simple extension of our algorithm applies to the
heterogeneous case where $\theta$ in~\eqref{eq:MRA:model} is randomly drawn from a finite family of linearly independent vectors. Using ideas initiated in the present paper, follow-up work~\cite{BanBluPer17} has confirmed that similar phenomena arise for molecule reconstruction in cryo-EM, albeit in a slightly weaker sense than the one presented in this paper.

\section{Overview}
In this section, we give an overview of our contributions and how they fit in the existing literature.

\subsection{Existing methods}

The difficulty of the multi-reference alignment problem resides in the fact that both the signal $\theta \in \R^d$ and the shifts $\ell_1,\ldots,\ell_n\in\ZZ_d$ are unknown. If the shifts were known, one could easily estimate $\theta$ by taking the average of $R_{\ell_i}^{-1}y_i, i =1, \ldots, n$. In fact, this simple observation is the basis of the so-called ``synchronization'' approach~\cite{ASinger_2011_angsync,Bandeira_Charikar_Singer_Zhu_Alignment,Bandeira_NonUniqueGames}: first estimate the shifts by $\tilde \ell_1,\ldots,\tilde \ell_n\in\ZZ_d$ and then estimate $\theta$ by averaging the $R_{\tilde \ell_i}^{-1}y_i$'s. While the synchronization approach can be employed at high $\SNR$, it is limited by the fact that at low $\SNR$, even alignment of observations to the true signal yields inaccurate shift estimates~\cite{Sapiro_limitsimagealignment}.

Instead, we take a different approach that exploits the connection between MRA and Gaussian mixture models. This connection is based on the fact that in MRA, the data $y$ is distributed according to a uniform mixture of Gaussians whose centers are the rotated vectors $R_1 \theta, \dots, R_{d} \theta$. To analyze MRA, we therefore rely on techniques from the Gaussian mixture model literature. One insight from this literature, which is crucial to our work, is that there are two separate estimation problems that can be posed for Gaussian mixture models. The first is \emph{clustering}, in which the goal is to assign a label to each datapoint corresponding to the Gaussian from which it was drawn. The second is \emph{parameter estimation}, in which the goal is simply to learn the Gaussians themselves---i.e., to identify the mean vectors and covariance matrices of each Gaussian component, without necessarily assigning a label to each point. Previous theoretical work on MRA has focused on the first task, which forms the basis for the synchronization approach. By contrast, our approach is based on the second task: we seek only to estimate the underlying parameters of the mixture; as they correspond to the underlying signal of interest. This connection also motivates our theoretical approach: we develop an approach based on the \emph{method of moments}, which was introduced in Pearson's seminal paper on Gaussian mixture models~\cite{Pea94} and has recently led to efficient estimators with provably optimal guarantees~\cite{MoiVal10}.

\subsection{The method of invariants}
In this work, we develop a new approach to MRA based on the method of moments.
This method focuses on the tensors $T^{(r)}(\theta)$ defined by
\begin{equation}\label{eq:moment_tensor}
 T^{(r)}(\theta) := \frac{1}{d}\sum_{\ell=1}^d   (R_\ell \theta)^{\otimes r} \,.
\end{equation}
These tensors are precisely the moments of the uniform distribution over the set of vectors $\{R_1 \theta, \dots, R_d \theta\}$.
We first establish that the parameter $\theta$ can be identified by the moment tensors $\{T^{(r)}(\theta)\}_{r \geq 1}$.
We then show that we can estimate the moment tensors accurately enough to recover the original signal.

The method of moments has an alternate interpretation in the context of MRA and similar problems involving group actions.
One striking fact is that the moment tensors in MRA capture features of the signal that are \emph{invariant} under cyclic shifts.
For example, the first moment tensor reduces to the entrywise mean of the signal (i.e., the vector in $\RR^d$ each of whose entries is the average value of $\theta$), which is an example of an invariant feature: it is clearly invariant under cyclic shifts of $\theta$ and, as we show below, can easily be estimated consistently in the MRA model.
More generally, each entry in the moment tensor $T^{(r)}(\theta)$ is an invariant polynomial in the coordinates of $\theta$, and these invariant polynomials can always be estimated in the MRA model as long as $\sigma$ is known.
We therefore call our approach the \emph{method of invariants}.

In what follows, we focus on the moment tensors $T^{(r)}(\theta)$ for $r \leq 3$.
Our core contribution is to show that estimation on the basis of these first three moment tensors yields optimal sample complexity as a function of the SNR for the MRA model.
We stress, however, that the focus on moment tensors is not limited to MRA, and that the method of invariants can be used to obtain sample complexity bounds for a wide variety of similar models.
In this work, we specialize to MRA since it provides perhaps the simplest nontrivial application of these ideas.

To state our results, we recall the definition of the Fourier transform.
Given a vector $\theta \in \RR^d$, we denote by $\hat \theta$ its Fourier transform, that is, the vector with entries \begin{equation*}\hat \theta_k  = \frac 1 d\sum_{j=1}^d e^{2 \pi i \frac{jk}{d}} \theta_j\,.\end{equation*}
This linear function which takes vectors to their Fourier transform extends naturally to a linear function on tensors: given an order-$r$ tensor $T$, we define its Fourier transform by \begin{equation*}\hat T_{a_1 \dots a_r} = \frac{1}{d^r} \sum_{1 \leq t_1, \dots, t_r \leq d} e^{2 \pi i \frac{a_1 t_1 + \dots + a_r t_r}{d}} T_{t_1 \dots t_r}\,.\end{equation*}

The first three moment tensors in the MRA model correspond to quantities often studied under different names in the signal processing literature.
In addition to the first moment tensor---which, as noted above, reduces to the entrywise mean of the signal---the second and third moment tensors are also easy to describe in terms of $\theta$.
The second moment tensor $T_2(\theta)$ corresponds to the autocorrelation of the signal $\theta$.
The Fourier transform of this tensor yields the power spectrum of $\theta$ (the square of the absolute value of the Fourier coefficients of the signal), which is often used as an invariant feature in signal processing.
Note that, in general, this quantity does not
carry enough information to allow for estimation of $\theta$, since it provides only the magnitudes of the Fourier coefficients, but not their phases.

The crucial object in the case of MRA is the third moment tensor.
The Fourier transform of this object is known as the \emph{bispectrum} of the signal, given by
\[
\BBB\left(k_1,k_2\right) = \hat \theta_{k_1}\hat \theta_{k_2}\hat \theta_{-k_1 - k_2},
\]
where $\hat \theta$ is the Fourier transform of $\theta$, $k_1,k_2\in [d]$, and the indices are taken modulo $d$.
The bispectrum was originally introduced in a statistical context~\cite{bispectrum_Brillinger,bispectrum_Tukey}, and it is known~\cite{kakarala2009completeness} that the bispectrum uniquely determines the signal $\theta$ up to cyclic shift whenever $\hat \theta_k \neq 0$ for all $k \in [d]$. We call such signals \emph{generic}, since this property is satisfied for all $\theta \in \RR^d$ apart from a set of measure zero.
In other words, for generic signals, the moment tensors $T_1(\theta)$, $T_2(\theta)$, and $T^{(3)}(\theta)$ suffice to identify the true signal $\theta$.
This fact has been exploited before to obtain estimates for alignment problems~\cite{bispectrum_Sadler,bispectrum_Giannakis, BenBouMa17}.

Note that the sample average based estimator for $T^{(3)}$ has a variance of order $\sigma^6/n$ when $\sigma$ is large, since it is a cubic polynomial of noisy data. It suggests that in the low $\SNR$ regime, any approach relying on the bispectrum requires at least order $1/\SNR^3$ samples. Since this dependency on $\SNR$ is very different from the $1/\SNR$ sample complexity of many models, bispectrum approaches seem highly suboptimal.

The main contribution of our work is to show that this number of samples is in fact a \emph{fundamental} requirement of the problem when the shifts are sampled from the uniform distribution, independent of the approach taken (following ideas developed in~\cite{BandeiraRigolletWeed17_MRA}): all estimators suffer from the same limitations, including the maximum-likelihood estimator (see Figure~\ref{fig:fred}). This shows that the latent cyclic transformations fundamentally change the difficulty of the problem. A similar phenomenon has been demonstrated for a Boolean version of MRA~\cite{AbePerSin17}.

To complement our lower bound, we also propose simple algorithm based on the method of moments capable of provably achieving the optimal $1/\SNR^3$ sample complexity for generic signals.
While other algorithms employing the bispectrum exist in the literature~\cite{bispectrum_Sadler,bispectrum_Giannakis, BenBouMa17}, ours has the virtue of acting directly to decompose the third moment tensor via a straightforward and principled approach.
As we note below, this simple algorithm also extends to the heterogenous setting, for which no algorithms enjoyed theoretical guarantees prior to this work.

\subsection{Non-generic signals}

The bispectrum-based methods for the multi-reference alignment problem we present work only for generic signals. In fact, non-generic signals can exhibit significantly worse behavior. Indeed, for non-generic signals, it can be shown that \emph{no} approach can succeed with fewer than $1/\SNR^{d}$ samples.
Nevertheless, this pessimistic scenario does not seem to be representative of signals encountered in practice. In fact, these signals that are hard to estimate form a set of zero Lebesgue measure. See~\cite{BandeiraRigolletWeed17_MRA} for more details.

\subsection{The heterogeneity problem}

One of the main challenges in cryo-EM reconstruction is the heterogeneity problem, where the noisy images one observes represent \emph{different} three-dimensional molecular structures. This problem arises often in practice, since even when a sample is perfectly purified, many large organic molecules naturally adopt different shapes, or conformations, depending on their environment and function.
A single sample can therefore include many different structures~\cite{scheres2016processing}.

The MRA model can be extended to accommodate heterogeneity by assuming that  in~\eqref{eq:MRA:model}, the vector $\theta \in \RR^d$ is also a latent variable drawn from a finite set of unknown vectors $\mathcal{C}=\{\theta^{(1)}, \ldots, \theta^{(K)}\}$. The goal here is to recover the set $\mathcal{C}$ up to a cyclic shift and the proportion of each $\theta^{(j)}$.

Our approach based on the method of invariants coupled with tensor decomposition techniques extends to the heterogeneous setup. It yields the first algorithm capable of provably solving the heterogeneous MRA at arbitrarily low $\SNR$, albeit at a potentially suboptimal sample complexity of $1/\SNR^{5}$.

\subsection{Connections to cryo-EM and XFEL}
One of the main motivations to study the multi-reference alignment problem is that it serves as a simpler surrogate for cryo-EM. This paper indicates potentially fruitful directions for future work. Our results offer theoretical support for the use of invariant methods in cryo-EM, a proposal which dates back to Zvi Kam~\cite{kam1980}.
These methods have also proven effective in XFEL structure determination~\cite{ArdMecGru18,DonSetZwa17}.

Our work serves as a first step towards a complete statistical theory of cryo-EM. In fact, follow-up work to this paper has demonstrated that the method of invariants can be used to characterize the sample complexity of more general models, including the reconstruction problem for cryo-EM~\cite{BanBluPer17}.

\subsection{Notation}
We use $[d]$ to represent the set $\{1, \dots, d\}$ and $I_d$ to represent the $d \times d$ identity matrix.
The smallest and largest singular values of a matrix are denoted $\sigma_{\min}$ and $\sigma_{\max}$, respectively. The symbol $\poly(\cdot)$ refers to an unspecified polynomial with constant coefficients. $C_d$ is used to refer to a constant that may depend on $d$ but not on other parameters, and it may refer to a different constant in different appearances throughout the text.
The expression $f(n) = O(g(n))$ means that there exists a constant $C$ such that $f(n) \leq C g(n)$ for all $n$, and we write $O_d(g(n))$ when the constant may depend on $d$. We write $g(n) = \Omega(f(n))$ when $f(n) = O(g(n))$.

\section{Fundamental limitations}
In this section, we establish the fundamental limits of MRA and point to shortcomings of existing strategies to achieve optimal sample complexity.

\subsection{Lower bounds for sample complexity}
Since observations in the MRA model~\eqref{eq:MRA:model} are invariant under a global cyclic shift, one may only identify $\theta$ up to such a global shift. To account for this fact, it is natural to employ the following shift-invariant distance between vectors $\theta, \tau \in \R^d$: 
\begin{equation*}
\metric(\theta, \tau) = \min_{\ell \in \ZZ_d} \|\theta - R_\ell \tau\|_2\,.
\end{equation*}

As noted in Section~\ref{sec:intro} above, by applying an independent and uniform random cyclic shift to each observation, we can always reduce the MRA model to the case where  $\ell_1, \ldots, \ell_n$ are drawn i.i.d.\ uniformly from $[d]$. In this case, the distribution of $y$ in~\eqref{eq:MRA:model} is a uniform mixture of the $d$ Gaussian distributions $\NNN(\theta, \sigma^2 I_d), \dots, \NNN(R_{d-1}\theta, \sigma^2 I_d)$.
If $y$ is generated according to this distribution, we call it a ``sample from MRA with signal $\theta$.''
The statistical properties of this Gaussian mixture are analyzed in~\cite{BandeiraRigolletWeed17_MRA}.

If $\sigma$ is small---that is, if the $\SNR$ is sufficiently large---then the signals can be aligned (for example, via the synchronization approach~\cite{ASinger_2011_angsync}), and therefore $\theta$ can be estimated accurately on the basis of $n$ samples from MRA with signal $\theta$ as long as $n \geq C /\SNR$ for some constant $C$.
This is the same dependence that would be expected in the absence of shifts.
Strikingly, the situation in the high-noise regime (when the $\SNR$ is low) is very different: estimation is impossible unless $n \geq C /\SNR^3$ for some constant $C$.

We recall that we call a signal \emph{generic} if $\hat \theta_k \neq 0$ for all $k \in [d]$.
\begin{theorem}\label{Thm:lowerbound_homogeneous}
Fix $d > 2$, $\eps > 0$ sufficiently small, and $\sigma \ge 1$. There exists a universal constant $C$ and a constant $c_d$ depending on $d$ such that, for any estimator $\tilde \theta$ based on $n$ samples from~\eqref{eq:MRA:model}, there exists with probability at least $1/4$ a generic signal $\theta \in \RR^d$ with $\|\theta\|_2 = 1$ and $|\hat \theta_k| \geq c_d > 0$ for all $k \in [d]$, such that $\rho(\tilde \theta, \theta) \ge \varepsilon$ when  $n \leq C \sigma^6\eps^{-2}$.
\end{theorem}
In other words, if we require that our estimator $\tilde \theta$ satisfy $\rho(\tilde \theta, \theta) < \eps$ with probability close to $1$, then we must have $n \geq C \sigma^6 \eps^{-2}$.
We prove this fact in the appendix using the tight information-theoretic bounds developed in~\cite{BandeiraRigolletWeed17_MRA}, which are based on the method of invariants and, in particular, on the observation given above that the second moment tensor does not carry enough information about $\theta$ in general.

Theorem~\ref{Thm:lowerbound_homogeneous} holds in the ``high-noise regime'' when $\sigma \geq 1$, and the synchronization approach~\cite{ASinger_2011_angsync} can be shown to succeed when $\sigma$ is sufficiently small. We leave the question of exploring the boundary between these regimes---including the question of whether there exists a sharp ``phase transition''---to future work.

\subsection{The importance of high frequencies}\label{sec:high}
As noted above, the sample complexity exhibited by the method of invariants, $1/\SNR^3$, is tight for generic signals.
For non-generic signals, while the method of invariants still yields optimal results (see~\cite{BanBluPer17}), the precise sample complexity depends on specific properties of the support of the Fourier transform of the original signal. As we we illustrate in this section, this dependence is often counter-intuitive.

Some approaches to the alignment problem implicitly adopt a strategy of first estimating low frequencies of a signal, and then using this initial estimate to estimate higher frequencies (see~\cite{barnett2016rapid}). In other words, these strategies assume that estimating a low-pass version of a signal is no harder than estimating the original signal.

\begin{figure}
\begin{center}
\includegraphics[width=0.9\textwidth]{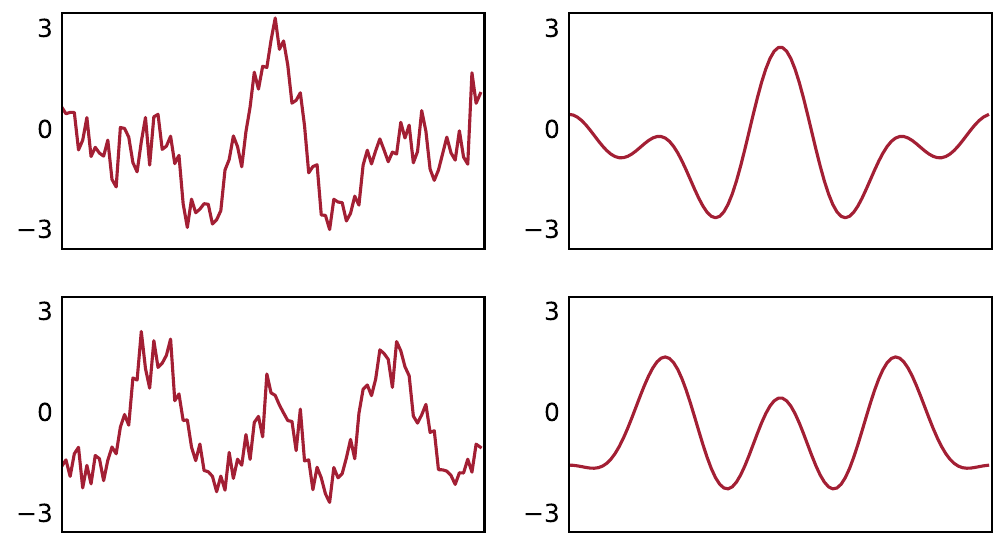}
\caption{Two signals whose Fourier transforms have almost full support (left column) and their corresponding low-pass versions (right column). Estimating either of the original signals is possible with $O_d(1/\SNR^3)$ samples. However, the same task for the low-pass versions requires $\Omega(1/\SNR^4)$ samples; in fact, even distinguishing between the two options requires this number of samples. This illustrates the importance of high frequencies in the MRA model.}\label{figure:example_lowpass}
\end{center}
\end{figure}

Surprisingly, this is \emph{not} the case in general, as following example shows.
Let us take $d \geq 14$ congruent to $2 \pmod{4}$ and $\theta\in\RR^d$ a signal whose Fourier transform $\hat \theta$ satisfies $\hat \theta_1 = \hat \theta_{-1} = 0$ but otherwise has full support.
We show in the appendix that we can estimate $\theta$ with $O_d\left(1/\SNR^3\right)$ samples because $\theta$ can be uniquely recovered from its bispectrum.
Surprisingly, if we low-pass $\theta$ by setting $\hat \theta_j = 0$ for all $|j| > 4$, then $\Omega\left(1/\SNR^4\right)$ samples are needed.
The difficulty in recovering the low-pass version arises from the following simple observation: if $\hat \theta_j = 0$ for all $j \notin \{\pm 2, \pm 3, \pm 4\}$, then the only nonzero entry of the bispectrum is $\BBB(2, 2)$. This implies that the bispectrum carries no information about the phase of $\hat \theta^{(3)}$. As we show in the appendix, this implies that $\Omega\left(1/\SNR^4\right)$ samples are required.
We illustrate this example in Figure~\ref{figure:example_lowpass}.

\section{Efficient recovery via tensor methods}\label{section:efficient}
Theorem~\ref{Thm:lowerbound_homogeneous} implies that the sample complexity of MRA for generic signals is at least $1/\SNR^3$ in the low-SNR regime.
In this section, we describe how the method of invariants also yields an efficient algorithm that achieves this optimal sample complexity, that is, it outputs an estimator $\tilde \theta$ of $\theta$ such that $\rho(\tilde \theta, \theta) \le \varepsilon$  with high probability whenever  $n \geq C_d \sigma^6\eps^{-2}$.

Our approach uses the method of invariants by estimating invariant features in  the third moment tensor $T^{(3)}$ defined in~\eqref{eq:moment_tensor}.
While other algorithms in the literature have also been based on recovering the signal on the basis of the third moment tensor via iterative methods~\cite{bispectrum_Sadler,bispectrum_Giannakis, BenBouMa17}, we propose a simpler procedure which also yields stable recovery guarantees.

First, we estimate $T^{(3)}$ by the following empirical quantity:
\begin{equation}
\label{eq:tildet3}
\tilde {T}^{(3)}_n = \frac{1}{dn} \sum_{i=1}^n \sum_{\ell=1}^d ((R_\ell y_i)^{\otimes 3} - 3 \sym(R_\ell y_i \otimes I_d))
\end{equation}
where
\begin{equation}
\label{eq:sym}
    \sym(A)_{a_1\dots a_r} = \frac{1}{r!} \sum_{\pi \in \cS_r} A_{\pi(a_1)\dots \pi(a_r)}\,.
\end{equation}
\begin{lemma}\label{lemma_estimating_T3}
The estimator $\tilde {T}^{(3)}$ is an unbiased estimator of $T^{(3)}$. Moreover, each entry of $\tilde T^{(3)}_n$ has a variance of order  $\sigma^6/n$ so long as $\sigma \geq 1$.
\end{lemma}
\begin{proof}
If $\xi_i \sim \cN(0, I_d)$, then both $\E[\xi_i]$ and $\E[\xi_i^{\otimes 3}]$ are zero.
This implies that%
\begin{align*}
    \frac 1 d \sum_{\ell=1}^d \E [(R_\ell y_i)^{\otimes 3}] &= \frac 1 d \sum_{\ell=1}^d \E[(R_\ell \theta + \sigma \xi)^{\otimes 3}] = \frac 1 d \sum_{\ell=1}^d ((R_\ell \theta)^{\otimes 3} + 3 \sym(( R_\ell \theta) \otimes I_d))\,,
\end{align*}
so $\tilde T^{(3)}_n$ is an unbiased estimator of $T^{(3)}$.

Each entry of $y_i$ is a Gaussian with variance $\sigma^2$, so the entries of $\sym(y_i \otimes I_d)$ have variance of order $\sigma^2$, and the entries of $y_i^{\otimes 3}$ have variance of order $\sigma^6$; the latter dominates for $\sigma \geq 1$. The claim follows.
\end{proof}
Then, we apply a basic decomposition technique, given in the next section, to the tensor $\tilde T^{(3)}_n$ to find a vector $\tilde \theta$ such that
\begin{equation*}
\tilde T^{(3)}_n \approx \frac 1d \sum_{\ell = 1}^d (R_\ell \tilde \theta)^{\otimes 3}\,.
\end{equation*}
The vector $\tilde \theta$ then serves as our estimate of $\theta$.

\subsection{Jennrich's Algorithm for Tensor Decomposition}
In this section, we detail a simple decomposition algorithm for the third moment tensor, which in turn provides an efficient algorithm that provably solves MRA for generic signals while achieving optimal sample complexity in terms of $\SNR$.
It involves the spectral decomposition of the tensor of empirical third moments. %
Such decompositions have been long studied and a sophisticated machinery has been developed over the years; see \cite[Chapter~3]{Moi17}.

The specific algorithm that we use is a standard tensor decomposition algorithm known as \emph{Jennrich's algorithm} (proposed in~\cite{harshman1970foundations} and credited to Robert Jennrich).
The version described below allows the recovery of vectors $u_1, \ldots, u_r$ (up to simple transformations) from a noisy version of the tensor
\begin{equation}
\label{eq:lowranktensor}
T = \sum_{j=1}^r u_j \otimes u_j \otimes v_j \in \R^{m \times m \times p},
\end{equation}
where $v_1, \ldots, v_r \in \R^p$ are arbitrary nonzero vectors.

\vspace{1ex}

{\centering\fbox{\parbox{.97\columnwidth}{
{\bf Jennrich's Algorithm} (\cite{harshman1970foundations,leurgans1993decomposition}).

\noindent{\sf Input:} Tensor $T \approx \sum_{j=1}^r u_j \otimes u_j \otimes v_j \in \R^{m \times m \times p}$.

\noindent{\sf Output:} Matrix $U=[\hat u_1, \dots, \hat u_r] \in \R^{m\times r}$%
\begin{itemize}
\item[$\blacktriangleright$] Choose random unit vectors $a, b \in \R^p$, and form matrices $A, B \in \R^{m\times m}$ with entries:
\[
A_{ij} = \sum_k T_{ijk} a_k, \ A=\sum_{j=1}^r \langle v_j, a\rangle u_j \otimes u_j,
\]
\[
B_{ij} = \sum_k T_{ijk}b_k,\ B=\sum_{j=1}^r \langle v_j, b\rangle u_j \otimes u_j
\]
\item[$\blacktriangleright$] Let $W$ be the matrix whose columns are the first $r$ left singular vectors of $A$.
\item[$\blacktriangleright$] Compute $M = W^\top A W (W^\top B W)^{-1}$.
\item[$\blacktriangleright$] Output $U = W P$, where $M = P D P^{-1}$ is the eigendecomposition of $M$.
\end{itemize}
}}}

\smallskip

Jennrich's algorithm requires only basic matrix operations and can therefore be implemented very efficiently even on large scale problems. It also enjoys the following robustness guarantees. Using the notation of Jennrich's algorithm, it is easy to see that $T^{(3)}$ is indeed a low-rank tensor of the form~\eqref{eq:lowranktensor}, with $m=p=d$, $u_j=v_j=R_{j-1}\theta$ (for $j=1, \ldots, r$) and $U = [\theta, R_{1}\theta, \dots, R_{d-1}\theta]$. 
We recall the following recovery guarantee of Jennrich's algorithm when applied to a tensor $\tilde T$ that is close to a low rank tensor.

\begin{theorem}[\cite{GoyVemXia14}, Theorem 5.2]\label{thm:jennrich-robust}
Let $T$ be a tensor of the form~\eqref{eq:lowranktensor} with all $u_j$ linearly independent, and define $\kappa(U) = \sigma_{\max}(U)/\sigma_{\min}(U)$. Moreover, fix $\eps>0$ and let $\tilde T$ satisfy $\|\tilde T - T\|_F \le \eps$. Then Jennrich's algorithm applied to $\tilde  T$ returns unit vectors $\tilde u_j, j=1, \ldots, r$ such that there exists a permutation $\pi$ and scalars $\beta_j$ satisfying
\begin{equation}\label{eq:jennrichestimator:sigma3_0}
\max_{j \in [r]}\| \tilde u_j - \beta_j u_{\pi(j)} \|_\infty \le \eps \poly(m,\kappa)
\end{equation}
with high probability.
\end{theorem}

Let $\tilde u_1$ be the first vector output by Jennrich's algorihtm applied to $\tilde T^{(3)}_n$ and let
$$
\tilde \beta_1 = \tilde u_1^\top \one /\tilde \mu\,, \quad \tilde \mu=\frac1n \sum_{i=1}^n y_i^\top \one\,, \quad \tilde \theta=\tilde u_1/\tilde \beta_1
$$

In the appendix, we show that an algorithm {\sf homoJen} based on Jennrich's algorithm for tensor decomposition applied to the $\tilde T^{(3)}_n$ enjoys the following theoretical guarantees.

\begin{theorem}\label{thm:homogeneous}
Fix $\sigma>.1$ and $\delta \in (0,1)$ and assume  $.1 \leq \|\theta\|_2 \leq 10$. Then, for any $\eps>0$ Jennrich's algorithm applied to $\tilde T^{(3)}_n$  outputs $\tilde \theta_n$ such that $\rho(\tilde \theta_n, \theta) \le \eps$ with probability at least $1-\delta$ whenever $$n \ge \sigma^{6} \eps^{-2} \poly(d,1/\min_{j \in [d]} |\hat \theta_j|,1/\delta)$$
in time $O(n d^3 + d^3 \poly(\log(1/\eps)))$.
\end{theorem}
Note that the constants $.1$ and $10$ are arbitrary and may be replaced by any other constants.
The time complexity is dominated by the time necessary to construct the empirical tensor $\tilde T^{(3)}_n$, which requires only a single pass over the data.
Since Jennrich's algorithm relies on basic matrix operations, it requires only $O(d^3 \poly(\log(1/\eps)))$ additional computation time once $\tilde T^{(3)}_n$ has been constructed.

In view of the lower bound appearing in Theorem~\ref{Thm:lowerbound_homogeneous}, the sample complexity of the modified Jennrich algorithm is optimal in terms of the SNR for all signals $\theta$ whose Fourier transform satisfies $|\hat \theta_j| \geq c > 0$ for all $j \in [d]$. On the other hand, Theorem~\ref{thm:homogeneous} gives no guarantee for non-generic signals whose Fourier transform is allowed to approach 0. This limitation is unavoidable: as mentioned above, \emph{no} approach based on the bispectrum can succeed for general signals. We refer the reader to \cite{BandeiraRigolletWeed17_MRA} for more details.

Several other bispectrum-based algorithms have appeared in the literature; see \cite{BenBouMa17} for a recent empirical study. These may perform better in practice, but they largely do not come with the theoretical guarantees of the algorithm proposed here, and they do not yield efficient algorithms for the heterogenous case discussed below.

\section{Heterogeneity}\label{section:heterogeneous}
In this section, we sketch an extension of the previous results to the heterogenous multi-reference alignment problem.
We recall this model here for completeness. %
In heterogenous MRA, we observe 
\begin{equation}
\label{EQ:heteroMRA}
y_i=R_{\ell_i} \theta^{(Z_i)} + \sigma \xi_i \,, \quad i=1, \ldots, n\,,
\end{equation}
where $Z_1, \ldots, Z_n \in \{1, \ldots, K\}$ are i.i.d.\ latent variables such that $\Pr(Z_i=k)=\pi_k, k \in [K]$ that are independent of all other variables and $\theta^{(k)} \in \R^d, k \in [K]$ are unknown vectors.
The other variables are specified as in the homogeneous model~\eqref{eq:MRA:model}. The goal here is to recover the set of vectors $\theta^{(k)} \in \R^d, k=1, \ldots, K$ up to a cyclic shift and the probability mass function $\{\pi_k\}_{k\in [K]}$. 

The method of invariants described above can be extended to handle the heterogeneous model~\eqref{EQ:heteroMRA}. In this case our method proceeds by estimating the mixtures of signals from %
an unbiased estimator $\tilde T^{(5)}_n$ for the $5$-tensor 
$$
T^{(5)} = \sum_{k=1}^K \sum_{\ell=1}^d \frac{\pi_k}{d} \big(R_\ell \theta^{(k)}\big)^{\otimes 5}\,. 
$$ 
In our analysis of homogenous MRA, we noted that the moment tensors $T^{(1)}=T^{(1)}(\theta), T^{(2)}=T^{(2)}(\theta), T^{(3)}=T^{(3)}(\theta)$ uniquely determine $\theta$, as long as $\theta$ is generic. The method of invariants can also be applied to the heterogenous case to show that the moment tensors $T^{(1)}, \dots, T^{(5)}$ determine the vectors $\theta^{(1)}, \dots, \theta^{(K)}$ as long as the vectors satisfy a particular genericity condition. Our proof of this fact is \emph{algorithmic} in the sense that we exhibit an efficient algorithm, which can recover the vectors $\theta^{(1)}, \dots, \theta^{(K)}$ as long as the collection is suitably generic. The fact that the method of invariants can be extended to the heterogenous case supports the idea that it is a flexible, general approach to models of this kind.
This algorithm achieves sample complexity $1/\SNR^5$, whereas the optimal sample complexity for heterogenous MRA is known to be $1/\SNR^3$ in several settings, including when $\theta^{(1)}, \dots, \theta^{(K)}$ are drawn independently from $\cN(0, I/d)$~\cite{BanBluPer17,Wei18}.
Our algorithm therefore does not achieve optimal sample complexity in general; however, it is the first efficient algorithm for the heterogenous problem with provable guarantees.

We now sketch the basic idea of our approach.
Using manipulations similar to the ones arising in the proof of Lemma~\ref{lemma_estimating_T3}, it is not hard to show that $T^{(5)}$ can be rewritten as
$$
 T^{(5)} = \E[y_i^{\otimes 5}] - 10 \sigma^2 \sym(T^{(3)} \otimes I_d) - 15 \sigma^4 \sym(T_1 \otimes I_d^{\otimes 2})
$$
where $T_1 = \E[y_i]$ and $T^{(3)}$ is defined in~\eqref{eq:moment_tensor}. Therefore an unbiased estimator of $T^{(5)}$ is given by
$$
\tilde T^{(5)}_n = \frac{1}{n}\sum_{i=1}^n y_i^{\otimes 5} - 10 \sigma^2 \sym(\tilde T^{(3)}_n \otimes I_d) - 15 \sigma^4 \sym(\frac{1}{n}\sum_{i=1}^n y_i \otimes I_d^{\otimes 2})\,,
$$
where $\tilde T^{(3)}_n$ is given in~\eqref{eq:tildet3}. Moreover, each entry of $\tilde T^{(5)}_n$ has variance of order $\sigma^{10}/n$ so long as $\sigma$ is bounded below by a positive constant.

We propose a method that consists in applying Jennrich's algorithm to an appropriate flattening of $\tilde T^{(5)}_n$.  We call it {\sf heteroJen}.
It hinges on the following observation: the $5$-tensor $T^{(5)}$ can be flattened into a $3$-tensor of shape $d^2 \times d^2 \times d$ that admits the following low-rank decomposition:
$$ \sum_{k=1}^K \sum_{\ell=1}^d \frac{\pi_k}{d} (R_\ell \theta^{(k)})^{\otimes 2} \otimes (R_\ell\theta^{(k)})^{\otimes 2} \otimes (R_\ell \theta^{(k)})\,. $$

The algorithm {\sf heteroJen} then proceeds by plugging $\tilde T^{(5)}_n$ into the above flattening operation and then applying Jennrich's algorithm to the resulting $3$-tensor of shape $d^2 \times d^2 \times d$. Theorem~\ref{thm:jennrich-robust} implies that this procedure outputs vectors $\tilde u_i$, $1 \leq i \leq dK$ with the following guarantees: there exist scalars $\beta_i$ and a bijection $a \times b : [dK] \to [d] \times [K]$ satisfying 
$$
\|\tilde u_i - \beta_i (R_{a(i)} \theta^{(b(i))})^{\otimes 2}\|_\infty \leq C_K\frac{\sigma^5}{\sqrt{n}}\poly(d)\,,
$$
with high probability. %

We compute $\tilde v_i$ as the leading eigenvector of the $d \times d$ matrix $\tilde u_i$. Letting $\tilde V^{(3)}$ be the $d^3 \times d K$ matrix with columns $\tilde v_i^{\otimes 3}$, we estimate $\tilde \alpha \in \RR^{dK}$ as the least-squares solution to $\tilde V^{(3)} \tilde \alpha = \mathrm{vec}(T^{(3)})$.  The vectors $\tilde w_i := \tilde \alpha^{1/3} \tilde  v_i$ (with entrywise exponentiation) now comprise $dK$ redundant estimates to the $K$ original signals $\theta_k$; we remove this redundancy by clustering these $dK$ estimates according to the pseudometric $\rho_2(x, y) = \min_{1 \leq \ell \leq d} \|x^{\otimes 2} - (R_\ell y)^{\otimes 2}\|_2$. (We show in the proof of Theorem~\ref{thm:heterogeneous}, below, that this clustering can be accomplished by a simple thresholding scheme.)
Finally, the procedure {\sf heteroJen} returns one vector from each cluster.

The {\sf heteroJen} procedure enjoys the following theoretical guarantees that rely on the following condition number.

Let $U = [\mathrm{vec}((R_1 \theta^{(1)})^{\otimes 2})$,$\ldots$,$\mathrm{vec}((R_d \theta^{(K)})^{\otimes 2})]$, and denote the condition number of $U$ by $\kappa$. It can be shown that $\kappa$ is generically finite. Indeed, suppose we have some nonzero linear relation $0 = \sum_{k=1}^K \sum_{\ell=1}^d c_{k,\ell} (R_\ell \theta^{(k)}) (R_\ell \theta^{(k)})^\top$. Taking the Fourier transform of this matrix and examining the $a,b$ entry, we have $0 = \sum_{k=1}^K \widehat{(c_k)}_{a-b} \hat\theta_{a}^{(k)} \hat\theta_{b}^{(k)}$. Some $\widehat{(c_k)}_\alpha$ is nonzero, yielding a nontrivial linear relation among the autocorrelation vectors
$v_k$, $1 \leq k \leq K$, with $v_{k,j} = \hat\theta_{j}^{(k)} \hat\theta_{-\alpha-j}^{(k)}$. These vectors satisfy the symmetry $v_{k,j} = \overline{v_{k,-\alpha-j}}$, but are generic on this subspace, which has dimension at least $\lceil d/2 \rceil$. Hence generically no such relation exists, and the matrix $U$ has finite condition number.

\begin{theorem}\label{thm:heterogeneous}
Fix $\sigma>.1$ and $\delta \in (0,1)$. Assume that  $.1 \leq \|\theta^{(k)}\|_2 \leq 10$ for all $k\in [K]$ and that $K \leq \lceil d/2 \rceil$. Then, for any $\eps>0$, the {\sf heteroJen} applied to $T^{(5)}_n$ outputs $\{\tilde \theta^{(1)}_n, \ldots, \tilde \theta^{(K)}_n\}$ such that 
$$
\sum_k \min_j \rho(\tilde \theta^{(j)}_n,\theta^{(k)}) \le \eps\,,
$$
with probability at least $1-\delta$ whenever $$n \ge C_K\sigma^{10} \eps^{-2} \poly(d, \kappa ,1/\delta)$$ in time $O(n d^5 + d^6\poly(\log(1/\eps)))$.
\end{theorem}

To the best of our knowledge, {\sf heteroJen} is the first efficient method for heterogeneous MRA at low $\SNR$. As noted above, follow-up work has shown that similar method-of-moments approaches based on tensor decomposition can efficiently achieve $1/\SNR^3$ sample complexity under the assumption that the components are drawn independently from $\cN(0, I/d)$~\cite{Wei18}.

\section{Concluding remarks}

In this paper, we characterize the sample complexity of MRA, a first step towards a better statistical understanding of cryo-EM.
In particular, we show that \emph{any} estimator requires at least $1/\SNR^3$ samples at low $\SNR$.

We also present an algorithm based on the method of invariants that provably solves the MRA problem with optimal sample complexity at low $\SNR$. We further show that the approach can be adapted to heterogenous problems, an extension of particular importance in cryo-EM where different biological molecules or conformations are often imaged together. Our approach is the first to yield theoretical guarantees on any procedure for heterogenous MRA and opens the door to a broader application of the method of invariants.

While this work constitutes a first step towards a statistical theory of 3-D molecule reconstruction in cryo-EM, many questions remain open. Since an earlier version of this manuscript was available, follow-up work has shown that our approach can be extended to molecule reconstruction in cryo-EM~\cite{BanBluPer17} and to MRA with nonuniform shifts~\cite{Abbe_MRA_nonuniform}. These works establish that the method of invariants yields optimal sample complexity in a wide variety of settings.

\appendix
\section{Proof of lower bound}

\subsection{Proof of Theorem~\ref{Thm:lowerbound_homogeneous}}
In what follows, let $c_d$ and $C$ be constants (with $c_d$ depending on $d$) whose value may change from line to line.
Let $\theta \in \RR^d$ satisfy $\|\theta\|_2 = 1$ and $|\hat \theta_k| \gtrsim 1/\sqrt d$ for all $k \in [d]$, where $\hat \theta$ is the Fourier transform of $\theta$.
Define $\tau$ by setting
\begin{equation*}
    \hat \tau_k = \left\{\begin{array}{ll} e^{i \delta}\hat \theta_1 & \text{ if $k = 1$} \\
    e^{-i \delta}\hat \theta_{-1} & \text{ if $k = -1$} \\
    \hat \theta_k & \text{ otherwise,}
    \end{array}\right.
\end{equation*}
where $\delta = c_d \eps$ for some constant $c_d$ chosen so that $|\hat \tau_1 - \hat \theta_1| \geq 2\eps$ as long as $\eps$ is sufficiently small.
Note that $\tau$ also satisfies $|\hat \tau_k| \gtrsim 1/\sqrt d$ for all $k \in [d]$.
The inequality $\rho(\theta, \tau) \geq 2\eps$ holds for all $\eps$ sufficiently small.
To see this, note that if $\delta$ is sufficiently small, then $|1 - e^{i 2 \pi k/d}| \geq 2 |1 - e^{i \delta}|$ for all $1 \leq k \leq d-1$.
This implies that $|\hat \theta_1 - \hat \tau_1| = \min_{k \in [d]} |\hat \theta_1 - e^{i 2\pi k/d}\hat{\tau}_1| $, and hence for any shift $R$ we have $\|\theta - R \tau\|_2 \geq |\hat \theta_1 - \widehat{R \tau}_1| \geq \min_{k \in [d]} |\hat \theta_1 - e^{i 2\pi k/d}\hat{\tau}_1| \geq |\hat \theta_1 - \hat \tau_1| \geq 2 \eps $.

We now establish that no procedure can distinguish between MRA with signal $\theta$ and MRA with signal $\tau$ on the basis of $n$ samples if $n \leq C \sigma^6 \eps^{-2}$ with probability greater than $3/4$.
To prove this, we reproduce the following theorem~\cite[Theorem~9]{BandeiraRigolletWeed17_MRA}, whose proof we sketch for completeness.

\begin{theorem}\label{thm:moment_matching}
Assume $\sigma \geq 1$.
Let $\theta$ and $\tau$ be two mean-zero signals satisfying $\rho(\theta, \tau) \leq \eps$ and $T^{(r)}(\theta) = T^{(r)}(\tau)$ for $r < k$.
If $P_\theta$ and $P_\tau$ are the Gaussian mixtures corresponding to MRA with signals $\theta$ and $\tau$ respectively, then the Kullback-Leibler divergence $D(P_\theta \, \|\, P_\tau)$ satisfies
\begin{equation*}
D(P_\theta \, \|\, P_\tau) \leq C_k \sigma^{-2k} \eps^2
\end{equation*}
for some constant $C_k$ depending on $k$.
\end{theorem}
\begin{proof}{[Sketch]
Let $\phi(x)$ be the density of a $d$-dimensional Gaussian with covariance $\sigma^2 I_d$, and let $\phi_\theta$ and $\phi_\tau$ be the densities of $P_\theta$ and $P_\tau$, respectively.
Let $R$ be a uniformly distributed random cyclic shift.
The convexity of the exponential function implies
\begin{equation*}
\phi_\tau(x) = \E \phi(x - R \tau) \geq \phi(x) e^{-\|\tau\|_2^2/2 \sigma^2}\,.
\end{equation*}

The $\chi^2$ divergence $\chi^2(P_\theta \, \|\, P_\tau)$ between $P_\theta$ and $P_\tau$ then satisfies
\begin{align*}
\chi^2(P_\theta \, \|\, P_\tau)&:= \int \frac{(\phi_\theta(x) - \phi_\tau(x))^2}{\phi_\tau(x)} \mathrm{d}x \\
&\leq e^{\|\tau\|_2^2/2 \sigma^2} \int (e^{-\|\theta\|_2^2/2\sigma^2} \E e^{\frac{x^\top R \theta}{\sigma^2}} - e^{- \|\phi\|_2^2/2 \sigma^2} \E e^{\frac{x^\top R \tau}{\sigma^2}})^2 \phi(x)\mathrm{d}x\,.
\end{align*}
Expanding the square, collecting terms, and integrating with respect to $x$ yields
\begin{equation*}
\chi^2(P_\theta \, \|\, P_\tau) \leq e^{\|\tau\|_2^2/2 \sigma^2} \E [e^{(R' \theta)^\top R \theta/\sigma^2} - 2 e^{(R' \theta)^\top R \tau/\sigma^2} + e^{(R' \theta)^\top R \theta/\sigma^2}]\,,
\end{equation*}
where $R'$ is an independent copy of $R$.
Expanding this quantity as a Taylor series and applying Fubini's theorem to interchange summation and expectation yields
\begin{equation*}
\chi^2(P_\theta \, \|\, P_\tau) \leq e^{\|\tau\|_2^2/2 \sigma^2} \sum_{r \geq 1} \frac{\|T^{(r)}(\theta) - T^{(r)}(\tau)\|_{HS}^2}{\sigma^{2r} r!}\,,
\end{equation*}
where $\|\cdot\|_{HS}^2$ represents the Hilbert-Schmidt norm.
By~\cite[Lemma~B.12]{BandeiraRigolletWeed17_MRA}, for $\eps$ sufficiently small,
\begin{equation*}
\|T^{(r)}(\theta) - T^{(r)}(\tau)\|_{HS}^2 \leq 12 \cdot 2^r \eps^2\,.
\end{equation*}
Combining this with the assumption that $T^{(r)}(\theta) = T^{(r)}(\tau)$ for $r < k$ yields
\begin{equation*}
\chi^2(P_\theta \, \|\, P_\tau) \leq e^{\|\tau\|_2^2/2 \sigma^2} \eps^2 \sum_{r \geq k} \frac{12 \cdot 2^r}{\sigma^{2r} r!} = C_k \sigma^{-2k} \eps^2\,.
\end{equation*}
Since $D(P_\theta \, \|\, P_\tau) \leq \chi^2(P_\theta \, \|\, P_\tau)$~\cite[Lemma~2.7]{tsybakov}, the claim follows.}
\end{proof}

The two signals $\theta$ and $\tau$ we have constructed are easily shown to satisfy $T^{(1)}(\theta) = T^{(1)}(\tau)$ and $T^{(2)}(\theta) = T^{(2)}(\tau)$, since their means and power spectra (i.e., the moduli of their Fourier transforms) agree.

By applying Theorem~\ref{thm:moment_matching} and Pinsker's inequality, we obtain
\begin{equation*}
\mathrm{TV}(P_\theta^{\otimes n}, P_\tau^{\otimes n})^2 \leq \frac 1 2 D(P_\theta^{\otimes n}\,\|\, P_\tau^{\otimes n}) \leq \frac 1 2 C \sigma^{-6} \eps^2 n\,.
\end{equation*}
Therefore if $n \leq C^{-1} \sigma^6 \eps^{-2}$, \cite[Theorem~2.2]{tsybakov} implies that for any measurable function $\psi: \RR^{d \times n} \to \{\theta, \tau\}$ of the data $y_1, \dots, y_n$, it holds
\begin{equation*}
P_{\theta}^n(\psi(y_1, \dots, y_n) = \tau) + P_{\tau}^n(\psi(y_1, \dots, y_n) = \theta) \geq 1-\mathrm{TV}(P_\theta^{\otimes n}, P_\tau^{\otimes n}) \geq 1/2\,.
\end{equation*}

In other words, any hypothesis test $\psi$ must incur type-I and type-II error of at least $1/2$.
Via Le Cam's two-point testing argument~\cite{lecam1973convergence}, this fact implies that any estimator $\tilde \theta$ is bound to incur error $\eps$ with probability at least $1/4$, as claimed.
\epr

\section{Analysis of low-pass example in Section~\ref{sec:high}}

To show that $\theta$ can be recovered with $O_d(1/\SNR^3)$ samples, it suffices to show that the phases of the Fourier coefficients of $\theta$ can be reconstructed uniquely from its bispectrum. Given a complex number $z$, denote by $\arg(z)$ its phase. By applying a cyclic shift, we can assume without loss of generality that $\arg(\hat \theta_2) \in [0, 4\pi/d)$ and that $\arg(\hat \theta_3) \in [0, \pi)$. It is easy to check that the identity $2 \sum_{k = 1}^{(d-6)/4} \arg(\BBB(2, 2k)) + \arg(\BBB((d-2)/2, (d-2)/2)) = \frac d 2 \arg(\hat \theta_2)$
holds modulo $2 \pi$, and the assumption that $\arg(\hat \theta_2) \in [0, 4\pi/d)$ implies that the choice of $\arg(\hat \theta_2)$ is unique. This implies that all even-indexed phases can be recovered.
We also have the simple identity $\arg(\hat \theta_6) + \arg(\BBB(3, 3)) = 2 \arg(\hat \theta_3)$ modulo $2 \pi$, and the assumption that $\arg(\hat \theta_3) \in [0, \pi)$ implies that the choice is unique. Combined with the knowledge of $\arg(\hat \theta_2)$, this implies recoverability of all odd-indexed phases.

To show that the low-pass signals require $n \geq C \sigma^8$ samples, we simply note that the first three moment tensors of the low-pass signals agree. Theorem~\ref{thm:moment_matching} therefore implies that the Kullback-Leibler divergence between the relevant distributions is at most $C \sigma^{-8}$. The same argument given in the proof of Theorem~\ref{Thm:lowerbound_homogeneous} establishes that any test attempting to distinguish between the low-pass signals incurs type-I and type-II error of at least $1/2$ unless $n \geq C \sigma^8$.

\section{Proof of upper bounds}
\subsection{Proof of Theorem~\ref{thm:homogeneous}}
Write $\kappa(\theta)$ for $1/(\min_{j \in [d]} |\hat \theta_j|)$.
It can be shown that the condition number $\kappa(U)$ of $U$ satisfies $\kappa(U) \leq  \max_{j,k} \{|\hat\theta_j|/|\hat\theta_k||\} \leq \kappa(\theta)$.
Theorem~\ref{thm:jennrich-robust} implies that Jennrich's algorithm applied to $\tilde T^{(3)}_n$ outputs $\tilde u_1$ satisfying
$$
\|\tilde u_1 -\beta_1 R_j \theta\|_\infty\le \frac{\sigma^3 \poly(d)}{\sqrt{n}}, \quad c>0\,,
$$
with high probability for some $j \in [d]$ and some $\beta_j \in \R$.

\newcommand{\comment}[1]{}
\comment{
Suppose that we have run Jennrich with sufficiently many samples such that for some $i$, $\|\tilde u_1 - \beta_i R_i \theta\|_\infty \leq \alpha$. We will bound the total estimation error as follows:
\begin{align*}
    \|\tilde \beta^{-1} \tilde u - R_i \theta\|_2 &\leq \| \tilde\beta^{-1} \tilde u - \beta_i^{-1} \tilde u \|_2 + \|\beta_i^{-1} \tilde u - R_i \theta \|_2 \\
    &= |\tilde\beta^{-1} - \beta_i^{-1}| + |\beta_i|^{-1} \alpha \\
    &= |\beta_i|^{-1} \left(\alpha + \left| \frac{\beta}{\tilde \beta} - 1 \right| \right),
\end{align*}
and it suffices to control these terms. Firstly, comparing $\beta_i$ and $\|\theta\|_2^{-1}$, we have
\begin{align*}
    \left| \|\tilde u\|_2 - \| \beta_i R_i\theta \|_2 \right| &\leq \alpha, \\
    \left| \|\theta\|_2^{-1} - |\beta_i| \right| &\leq \alpha/\|\theta\|_2,
    \intertext{so that}
    \frac{1-\alpha}{\|\theta\|_2} &\leq |\beta_i| \leq \frac{1+\alpha}{\|\theta\|_2}.
\end{align*}

Next, by Chebyshev, we have $|\mu - \tilde \mu| \leq \frac{\sigma \sqrt{\delta}}{\sqrt{2 n d}}$ with probability $1 - \delta/2$. Thus comparing $\beta$ and $\tilde \beta$, we have
\begin{align*}
    \alpha &\geq \| \tilde u - \beta_i R_i \theta\|_\infty \left\| \frac{1}{d} \one \right\|_1 \\
    &\geq \left|\langle \tilde u - \beta_i R_i \theta, \frac1d \one \rangle\right| \\
    &= |\langle \tilde u, \one \rangle/d - \beta_i \mu|, 
\\%
    \alpha + |\beta_i| |\tilde \mu - \mu| &\geq |\langle \tilde u, \one \rangle/d - \beta_i \tilde \mu|, \\
    \langle \tilde u, \frac1d \one \rangle^{-1} \left( \alpha + |\beta_i| \frac{\sigma \sqrt{\delta}}{\sqrt{2nd}} \right) &\geq \left| \frac{\beta_i}{\tilde \beta} - 1 \right|.
\end{align*}
We know that $| \langle \tilde u, \frac1d \one \rangle - \beta_i \mu | \leq \alpha$, so $\langle \tilde u, \frac1d \one \rangle^{-1} \leq (|\beta_i \mu| - \alpha)^{-1}$. Note that $\sigma_{\min}(U) \leq \left\|U \cdot \frac{1}{\sqrt{d}} \one \right\|_2 = d|\mu|$, and also $\sigma_{\max}(U) \geq \|\theta\|_2$, so that $|\mu| \geq d \kappa(\theta) / \|\theta\|_2$; thus we have
$$ \left| \frac{\beta_i}{\tilde \beta} - 1 \right| \leq \left( |\beta_i| \frac{d \kappa(\theta)}{\|\theta\|_2} - \alpha \right)^{-1} \left( \alpha + |\beta_i| \frac{\sigma \sqrt{\delta}}{\sqrt{2nd}} \right). $$
Assembling, we have in total
\begin{align*}
    \rho(\tilde \beta^{-1} \tilde u, \theta) \leq \|\tilde \beta^{-1} \tilde u - R_i \theta \|_2 &\leq |\beta_i|^{-1} \left( \alpha + \left(|\beta_i| \frac{d \kappa(\theta)}{\|\theta\|_2} - \alpha \right)^{-1} \left( \alpha + |\beta_i| \frac{\sigma \sqrt{\delta}}{\sqrt{2nd}} \right) \right) \\
    &\leq |\beta_i|^{-1} \left( \alpha + \left(\frac{(1-\alpha) d \kappa(\theta)}{10^2} - \alpha \right)^{-1} \left( \alpha + 10 (1+\alpha) \frac{\sigma \sqrt{\delta}}{\sqrt{2nd}} \right) \right).
\end{align*}
}

As we are only concerned with polynomial dependence and not detailed bounds, we write $A \approx B$ if we can bound $|A - B| \leq \alpha \poly(d,\kappa(\theta),\delta^{-1}) + \sigma/\sqrt{n} \poly(d,\kappa(\theta),\delta^{-1})$ so long as\footnote{The precise bound of $1$ here is arbitrary.} $\alpha \leq 1$ and $\sigma/\sqrt{n} \leq 1$; we apply this also to vectors in $2$-norm or (equivalently) most other common norms.

Theorem~\ref{thm:jennrich-robust} guarantees us that $\tilde u \approx \beta_i R_i \theta$. Taking norms, we have $1 \approx |\beta_i| \|\theta\|_2$; as $\|\theta\|_2$ is bounded above and below by constants, we have that $|\beta_i| \approx 1/\|\theta\|_2$ is also of constant order. Note also that by Chebyshev we have that $|\tilde \mu - \mu| \leq \sigma \sqrt{\delta}/\sqrt{2nd}$ with probability $1-\delta/2$, so that $\mu \approx \tilde\mu$. From $\tilde u \approx \beta_i R_i \theta$ we also derive that
\begin{equation} \langle \tilde u,\one\rangle/d \approx \beta_i \mu \approx \beta_i \tilde\mu. \label{eq:pre-divide-muhat} \end{equation}
We know that $\|\theta\|_2 \leq \sigma_{\max}(U)$ and that $\sigma_{\min}(U) \leq \|U \cdot \frac{1}{\sqrt{d}} \one\|_2 = d|\mu|$, so that $|\tilde\mu| \approx |\mu| \geq d \kappa(\theta) / \|\theta\|_2$; we are thus justified in dividing \eqref{eq:pre-divide-muhat} by $\tilde\mu$ to obtain $\tilde \beta \approx \beta_i$, and $\beta_i / \tilde \beta \approx 1$. We now bound the total estimation error as follows:
\begin{align*}
    \|\tilde \beta^{-1} \tilde u - R_i \theta\|_2 &\leq \| \tilde\beta^{-1} \tilde u - \beta_i^{-1} \tilde u \|_2 + \|\beta_i^{-1} \tilde u - R_i \theta \|_2 \\
    &\leq |\tilde\beta^{-1} - \beta_i^{-1}| + |\beta_i|^{-1} \alpha \\
    &= |\beta_i|^{-1} \left(\alpha + \left| \frac{\beta}{\tilde \beta} - 1 \right| \right) \approx 0.
\end{align*}

Thus in order to bound this estimation error to within $\eps$, it suffices to require bounds of the form $\sigma / \sqrt{n} \leq \eps / \poly(d,\kappa(\theta),\delta^{-1})$ and $\alpha \leq \eps / \poly(d,\kappa(\theta),\delta^{-1})$. By Theorem~\ref{thm:jennrich-robust}, we achieve this bound on $\alpha$ from Jennrich's algorithm so long as $\|\tilde T^{(3)}_n - T^{(3)}\|_F \leq \eps / \poly(d,\kappa(\theta),\delta^{-1})$. This estimation error is achieved with probability $1 - \delta/2$ so long as $n \geq \sigma^6 \poly(d,\kappa(\theta),\delta^{-1})$, which also subsumes the explicit bound on $\sigma/\sqrt{n}$. By a union bound over the two probabilistic steps in this argument, the desired accuracy guarantee holds with probability $1-\delta$.
\epr

\subsection{Proof of Theorem~\ref{thm:heterogeneous}}

As in the proof of Theorem~\ref{thm:homogeneous}, we write $A \approx B$ if we can bound $$|A - B| \leq \alpha \poly(d,\kappa,\delta^{-1}) + \sigma^3/\sqrt{n} \poly(d,\kappa,\delta^{-1})$$ given that\footnote{The precise bound of $1$ here is arbitrary.} $\alpha \leq 1$ and $\sigma^3/\sqrt{n} \leq 1$; we apply this also to vectors in $2$-norm or (equivalently) most other common norms.

From Theorem~\ref{thm:jennrich-robust}, we are guaranteed that $\tilde u_i \approx \beta_i (R_{a(i)} \theta^{(b(i))})^{\otimes 2}$; taking norms, we have $1 \approx |\beta_i| \|\theta^{(b(i))}\|_2$, so that $|\beta_i| \approx \|\theta^{(b(i))}\|^{-1}$ is of constant order. Now by the Davis--Kahan theorem, if $v_{\max}(M)$ denotes either choice of unit-length eigenvector of $M$ corresponding to the eigenvalue of largest magnitude, we have
$$ \tilde v_i := v_{\max}(\tilde u_i) \approx \eps_i R_{a(i)} \theta^{(b(i))}/\|\theta^{(b(i))}\|_2, $$
for some sign $\eps_i = \pm 1$. Then we have $\tilde V^{(3)} \approx V^{(3)}$, where as above, $\tilde V^{(3)}$ is the $d^3 \times dK$ matrix whose columns are $\tilde v_i^{\otimes 3}$, and $V^{(3)}$ has columns $\eps_i (R_{a(i)} \theta^{(b(i))})^{\otimes 3}/\|\theta^{(b(i))}\|_2^3$. Estimating $T^{(3)}$ by $\tilde T^{(3)}_n$ according to Lemma~\ref{lemma_estimating_T3}, we have $\tilde T^{(3)}_n \approx T^{(3)}$ by Chebyshev, with probability $1-\delta/2$. Note then that $V^{(3)} \alpha = \mathrm{vec}(T^{(3)})$, where $\alpha_i = \eps_i \|\theta^{(b(i))}\|_2^3 / dK$. By the perturbation theory of linear systems, %
we are now guaranteed that, letting $\tilde \alpha$ be the least squares solution to $\tilde V^{(3)} \tilde\alpha = \mathrm{vec}(\tilde T^{(3)}_n)$, we have $\tilde\alpha \approx \alpha$, so long as the system is well-conditioned, which we defer to the following lemma:
\begin{lemma}\label{lemma:V3_condition} 
If $\kappa(V^{(3)})$ denotes the condition number of $V^{(3)}$, then
$\kappa(V^{(3)}) \leq \kappa \poly(d)$. \end{lemma}
As $\alpha_i$ is of constant order, it follows that $\tilde w_i := \tilde\alpha_i^{1/3} \tilde v_i \approx R_{a(i)} \theta^{(b(i))}$, so that $\rho(\tilde\alpha_i^{1/3} \tilde v_i, \theta^{(b(i))}) \approx 0$. We are thus guaranteed $dK$ good estimates to the original $K$ signals. We next discuss how to remove this redundancy by clustering.

Define the pseudometric on $\RR^d$ defined by $\rho_2(x,y) = \min_{1 \leq \ell \leq d} \| x^{\otimes 2} - (R_\ell y)^{\otimes 2} \|_2$. Note that
$$\rho_2(w_i,w_{i'}) \approx \rho_2(R_{a(i)} \theta^{(b(i))}, R_{a(i')} \theta^{(b(i'))}) = \rho_2(\theta^{(b(i))},\theta^{(b(i'))}). $$
If $b(i) = b(i')$, so that the two estimates $w_i$ and $w_{i'}$ should represent the same signal, we thus have $\rho(w_i,w_{i'}) \approx 0$. If $b(i) \neq b(i')$, we have
$$ \rho_2(w_i,w_{i'}) \approx \rho_2(\theta^{(b(i))},\theta^{(b(i'))}) = \min_\ell \| U (e_{b(i),0} - e_{b(i'),\ell}) \|_2 \geq \sqrt{2}\, \sigma_{\min}(U) = 1/\poly(d,\kappa), $$
where $e_{b,\ell} \in \RR^{dK}$ is the standard basis vector corresponding to signal $b$ and rotation $\ell$. It follows that, provided $\alpha$ and $\sigma^6/n$ are inverse-polynomially small in $d,\kappa,\delta^{-1}$, we exactly recover the clusters of estimates $\tilde w_i$ corresponding to the same signal $\theta_k$, simply by comparing on the metric $\rho_2$ and thresholding. Drawing one estimate $\tilde w_i$ from each cluster, we obtain one estimate of each signal.

To conclude, in order to bound this estimation error to within $\eps$, it suffices to require bounds of the form $\sigma^3 / \sqrt{n} \leq \eps / \poly(d,\kappa,\delta^{-1})$ and $\alpha \leq \eps / \poly(d,\kappa,\delta^{-1})$. By Theorem~\ref{thm:jennrich-robust}, we achieve this bound on $\alpha$ from Jennrich's algorithm so long as $$\|\tilde T^{(5)}_n - T^{(5)}\|_F \leq \eps / \poly(d,\kappa,\delta^{-1}).$$ This estimation error is achieved with probability $1 - \delta/2$ so long as $n \geq \sigma^{10} \poly(d,\kappa,\delta^{-1})$, which also subsumes the explicit bound on $\sigma^3/\sqrt{n}$. By a union bound over the two probabilistic steps in this argument, the desired accuracy guarantee holds with probability $1-\delta$.
\epr

\subsection{Proof of Lemma~\ref{lemma:V3_condition}}
We apply the following transformations which do not alter the condition number: we transform the rows by the third tensor power of a DFT, we permute the columns to sort by signal and rotation, and we negate columns according to the signs $\eps_i$. It thus suffices to control the condition number of the $d^3 \times dK$ matrix ${V^{(3)}}'$ whose columns are $(\hat R_j \hat{\theta_k})^{\otimes 3} / \|\theta_k\|_2^3$, where $\hat R_i = \mathrm{diag}(\{\omega^{ij}\}_j)$ is the Fourier representation of a rotation action ($\omega = e^{2 \mathrm{i} \pi / d}$), and $\hat \theta$ is the Fourier transform of $\theta$. Meanwhile, let $V_2'$ be the $d^2 \times dK$ matrix with columns $(\hat R_j \hat{\theta_k})^{\otimes 2}$, the Fourier transform of $U$, so that $\kappa(V_2') = \kappa$.

Let $v \in \RR^{dK}$; then we have
\begin{align*}
    \|{V^{(3)}}' v\|_2^2 &= \sum_{\ell=1}^d \left\| V_2' \mathrm{diag}\left(\{ \omega^{j\ell} (\hat\theta_k)_\ell \|\theta_k\|_2^{-3} \}_{jk} \right)\, v \right\|_2^2 \\
    &\geq \sum_{\ell=1}^d \sigma_{\min}(U)^2 \left\| \mathrm{diag}\left(\{ \omega^{j\ell} (\hat\theta_k)_\ell \|\theta_k\|_2^{-3} \}_{jk}\right)\, v \right\|_2^2 \\
    &= \sum_\ell \sigma_{\min}(U)^2 \sum_{jk} |(\tilde\theta_k)_\ell|^2 \|\theta_k\|_2^{-6} v_{jk}^2 \\
    &= \sigma_{\min}(U)^2 \sum_k \|\theta_k\|_2^{-4} \left(\sum_j v_{jk}^2 \right) \\
    &\geq \sigma_{\min}(U)^2 10^{-4} \|v\|_2^2,
\end{align*}
so that $\sigma_{\min}({V^{(3)}}') \geq \sigma_{\min}(U) 10^{-2}$. Observing the norms of columns, it is clear that $\sigma_{\max}(U)$ and $\sigma_{\max}({V^{(3)}}')$ are bounded above by $\poly(d)$, so we conclude that $\kappa(V^{(3)}) = \kappa({V^{(3)}}') \leq \kappa \poly(d)$, as desired.
\epr

\section*{Acknowledgments}
The authors thank Alex Wein for many insightful discussions on the topic of this paper.
\bibliographystyle{alphaabbr}
\bibliography{afonsomra}

\newcommand{\etalchar}[1]{$^{#1}$}
\begin{thebibliography}{BBSK{\etalchar{+}}18}

\bibitem[ABL{\etalchar{+}}17]{Abbe_MRA_nonuniform}
E.~Abbe, T.~Bendory, W.~Leeb, J.~Pereira, N.~Sharon, and A.~Singer.
\newblock Multireference alignment is easier with an aperiodic translation
  distribution.
\newblock {\em arXiv preprint arXiv:1710.02793}, 2017.

\bibitem[ADBS16]{Sapiro_limitsimagealignment}
C.~Aguerrebere, M.~Delbracio, A.~Bartesaghi, and G.~Sapiro.
\newblock Fundamental limits in multi-image alignment.
\newblock {\em IEEE Trans. Signal Process.}, 64(21):5707--5722, 2016.

\bibitem[APS17]{AbePerSin17}
E.~Abbe, J.~Pereira, and A.~Singer.
\newblock Sample complexity of the boolean multireference alignment problem.
\newblock In {\em 2017 IEEE International Symposium on Information Theory
  (ISIT)}, July 2017.

\bibitem[ASS{\etalchar{+}}09]{SAgarwal_etal_2009_Rome}
S.~Agarwal, N.~Snavely, I.~Simon, S.~M. Seitz, and R.~Szeliski.
\newblock Building rome in a day.
\newblock In {\em Twelfth IEEE International Conference on Computer Vision
  (ICCV 2009)}, Kyoto, Japan, September 2009. IEEE.

\bibitem[BBM{\etalchar{+}}17]{BenBouMa17}
T.~Bendory, N.~Boumal, C.~Ma, Z.~Zhao, and A.~Singer.
\newblock Bispectrum inversion with application to multireference alignment.
\newblock {\em Available online at arXiv:1705.00641 [cs.IT]}, 2017.

\bibitem[BBSK{\etalchar{+}}18]{BanBluPer17}
A.~S. Bandeira, B.~Blum-Smith, J.~Kileel, A.~Perry, J.~Weed, and A.~S. Wein.
\newblock Estimation under group actions: recovering orbits from invariants.
\newblock {\em arXiv preprint arXiv:1712.10163}, 2018.

\bibitem[BCS15]{Bandeira_NonUniqueGames}
A.~S. Bandeira, Y.~Chen, and A.~Singer.
\newblock Non-unique games over compact groups and orientation estimation in
  cryo-{EM}.
\newblock {\em Available online at arXiv:1505.03840 [cs.CV]}, 2015.

\bibitem[BCSZ14]{Bandeira_Charikar_Singer_Zhu_Alignment}
A.~S. Bandeira, M.~Charikar, A.~Singer, and A.~Zhu.
\newblock Multireference alignment using semidefinite programming.
\newblock In {\em I{TCS}'14---{P}roceedings of the 2014 {C}onference on
  {I}nnovations in {T}heoretical {C}omputer {S}cience}, pages 459--470. ACM,
  New York, 2014.

\bibitem[BGPS17]{barnett2016rapid}
A.~Barnett, L.~Greengard, A.~Pataki, and M.~Spivak.
\newblock Rapid solution of the cryo-{EM} reconstruction problem by frequency
  marching.
\newblock {\em SIAM J. Imaging Sci.}, 10(3):1170--1195, 2017.

\bibitem[Bri91]{bispectrum_Brillinger}
D.~R. Brillinger.
\newblock Some history of the study of higher-order moments and spectra.
\newblock {\em Statist. Sinica}, 1(2):465--476, 1991.

\bibitem[BRW17]{BandeiraRigolletWeed17_MRA}
A.~S. Bandeira, P.~Rigollet, and J.~Weed.
\newblock Optimal rates of estimation for multi-reference alignment.
\newblock {\em Available online at arXiv:1702.08546 [math.ST]}, 2017.

\bibitem[CBB{\etalchar{+}}06]{ChaBarBog06}
H.~N. Chapman, A.~Barty, M.~J. Bogan, S.~Boutet, M.~Frank, S.~P. Hau-Riege,
  S.~Marchesini, B.~W. Woods, S.~Bajt, W.~H. Benner, et~al.
\newblock Femtosecond diffractive imaging with a soft-x-ray free-electron
  laser.
\newblock {\em Nature Physics}, 2(12):839, 2006.

\bibitem[DM98]{dryden98_shape}
I.~L. Dryden and K.~V. Mardia.
\newblock {\em Statistical shape analysis}.
\newblock Wiley series in probability and statistics. Wiley, Chichester, 1998.

\bibitem[DSZ17]{DonSetZwa17}
J.~J. Donatelli, J.~A. Sethian, and P.~H. Zwart.
\newblock Reconstruction from limited single-particle diffraction data via
  simultaneous determination of state, orientation, intensity, and phase.
\newblock {\em Proceedings of the National Academy of Sciences},
  114(28):7222--7227, 2017.

\bibitem[Fra06]{Fra06}
J.~Frank.
\newblock {\em Three-dimensional electron microscopy of macromolecular
  assemblies: visualization of biological molecules in their native state}.
\newblock Oxford University Press, 2006.

\bibitem[FZB02]{foroosh02_subpixelregistration}
H.~Foroosh, J.~Zerubia, and M.~Berthod.
\newblock Extension of phase correlation to subpixel registration.
\newblock {\em {IEEE} Trans. Image Processing}, 11(3):188--200, 2002.

\bibitem[GC07]{GafCha07}
K.~Gaffney and H.~Chapman.
\newblock Imaging atomic structure and dynamics with ultrafast x-ray
  scattering.
\newblock {\em Science}, 316(5830):1444--1448, 2007.

\bibitem[Gia89]{bispectrum_Giannakis}
G.~B. Giannakis.
\newblock Signal reconstruction from multiple correlations: Frequency-and
  time-domain approaches.
\newblock {\em JOSA A}, 6(5):682--697, 1989.

\bibitem[GVX14]{GoyVemXia14}
N.~Goyal, S.~Vempala, and Y.~Xiao.
\newblock Fourier {PCA} and robust tensor decomposition.
\newblock In {\em Proceedings of the 46th Annual ACM Symposium on Theory of
  Computing}, pages 584--593. ACM, 2014.

\bibitem[Har70]{harshman1970foundations}
R.~Harshman.
\newblock Foundations of the {PARAFAC} procedure: Model and conditions for an
  explanatory multimodal factor analysis.
\newblock Technical report, Tech. Rep. UCLA Working Papers in Phonetics 16,
  University of California, Los Angeles, Los Angeles, CA, December. 13, 27,
  1970.

\bibitem[Kak09]{kakarala2009completeness}
R.~Kakarala.
\newblock Completeness of bispectrum on compact groups.
\newblock {\em arXiv preprint arXiv:0902.0196}, 2009.

\bibitem[Kam80]{kam1980}
Z.~Kam.
\newblock The reconstruction of structure from electron micrographs of randomly
  oriented particles.
\newblock {\em Journal of Theoretical Biology}, 82(1):15--39, 1980.

\bibitem[LC73]{lecam1973convergence}
L.~Le~Cam.
\newblock Convergence of estimates under dimensionality restrictions.
\newblock {\em The Annals of Statistics}, pages 38--53, 1973.

\bibitem[LRA93]{leurgans1993decomposition}
S.~Leurgans, R.~Ross, and R.~Abel.
\newblock A decomposition for three-way arrays.
\newblock {\em SIAM Journal on Matrix Analysis and Applications},
  14(4):1064--1083, 1993.

\bibitem[Moi14]{Moi17}
A.~Moitra.
\newblock Algorithmic aspects of machine learning.
\newblock {\em Lecture notes (MIT)}, 2014.

\bibitem[MV10]{MoiVal10}
A.~Moitra and G.~Valiant.
\newblock Settling the polynomial learnability of mixtures of gaussians.
\newblock In {\em 51th Annual {IEEE} Symposium on Foundations of Computer
  Science, {FOCS} 2010, October 23-26, 2010, Las Vegas, Nevada, {USA}}, pages
  93--102. {IEEE} Computer Society, 2010.

\bibitem[Pea94]{Pea94}
K.~Pearson.
\newblock Contributions to the mathematical theory of evolution.
\newblock {\em Philosophical Transactions of the Royal Society of London A:
  Mathematical, Physical and Engineering Sciences}, 185:71--110, 1894.

\bibitem[RCBL16]{Rosen_SLAM_PCC}
D.~Rosen, L.~Carlone, A.~Bandeira, and J.~Leonard.
\newblock A certifiably correct algorithm for synchronization over the special
  {Euclidean} group.
\newblock In {\em Intl. Workshop on the Algorithmic Foundations of Robotics
  (WAFR)}, San Francisco, CA, December 2016.

\bibitem[Sch16]{scheres2016processing}
S.~H. Scheres.
\newblock Processing of structurally heterogeneous cryo-{EM} data in {RELION}.
\newblock In {\em Methods in enzymology}, volume 579, pages 125--157. Elsevier,
  2016.

\bibitem[SG92]{bispectrum_Sadler}
B.~M. Sadler and G.~B. Giannakis.
\newblock Shift- and rotation-invariant object reconstruction using the
  bispectrum.
\newblock {\em Oct. Soc. Am. A}, 9:57--69, 1992.

\bibitem[Sig98]{Sig98}
F.~Sigworth.
\newblock A maximum-likelihood approach to single-particle image refinement.
\newblock {\em Journal of structural biology}, 122(3):328--339, 1998.

\bibitem[Sin11]{ASinger_2011_angsync}
A.~Singer.
\newblock Angular synchronization by eigenvectors and semidefinite programming.
\newblock {\em Appl. Comput. Harmon. Anal.}, 30(1):20 -- 36, 2011.

\bibitem[SSK13]{sonday2011_sim}
B.~Sonday, A.~Singer, and I.~G. Kevrekidis.
\newblock Noisy dynamic simulations in the presence of symmetry: Data alignment
  and model reduction.
\newblock {\em Computers \& Mathematics with Applications}, 65(10):1535 --
  1557, 2013.

\bibitem[SWZ{\etalchar{+}}18]{SchWanZha18}
J.~Schnitzbauer, Y.~Wang, S.~Zhao, M.~Bakalar, T.~Nuwal, B.~Chen, and B.~Huang.
\newblock Correlation analysis framework for localization-based superresolution
  microscopy.
\newblock {\em Proceedings of the National Academy of Sciences},
  115(13):3219--3224, 2018.

\bibitem[Tsy09]{tsybakov}
A.~B. Tsybakov.
\newblock {\em Introduction to nonparametric estimation}.
\newblock Springer Series in Statistics. Springer, New York, 2009.
\newblock Revised and extended from the 2004 French original, Translated by
  Vladimir Zaiats.

\bibitem[Tuk84]{bispectrum_Tukey}
J.~W. Tukey.
\newblock The spectral representation and transformation properties of the
  higher moments of stationary time series.
\newblock In D.~R. Brillinger, editor, {\em The Collected Works of John W.
  Tukey}, volume~1, chapter~4, pages 165--184. Wadsworth, 1984.

\bibitem[vAMG18]{ArdMecGru18}
B.~von Ardenne, M.~Mechelke, and H.~Grubm{\"u}ller.
\newblock Structure determination from single molecule x-ray scattering with
  three photons per image.
\newblock {\em Nature communications}, 9(1):2375, 2018.

\bibitem[Wei18]{Wei18}
A.~Wein.
\newblock {\em Statistical Estimation in the Presence of Group Actions}.
\newblock PhD thesis, MASSACHUSETTS INSTITUTE OF TECHNOLOGY, 2018.

\end{thebibliography}

\end{document}